\documentclass[preprint,12pt]{elsarticle}

\usepackage{amssymb}
\usepackage{amsmath}
\usepackage{subfigure}
\begin{document}

\begin{frontmatter}



\title{Mixed-bond spin-1 Ising model with nonlinear interactions for the Fe-Mn alloys
}


\author[label1]{A. S. Freitas}
\address[label1]{DFI, CCET, Universidade Federal de Sergipe, S\~ao Crist\'ov\~ao, 49100-000, SE, Brazil}

\author[label2]{Douglas F. de Albuquerque}
\address[label2]{DMA, CCET, Universidade Federal de Sergipe, S\~ao Crist\'ov\~ao, 49100-000, SE, Brazil}

\author[label3]{I. P. Fittipaldi}
\address[label3]{Representa\c{c}\~ao Regional do Minist\'erio da Ci\^encia, Tecnologia e Inova\c{c}\~ao no Nordeste - ReNE, 50740-540, Recife, PE, Brazil
}
\author{N. O. Moreno}
\address[label4]{DFI, CCET, Universidade Federal de Sergipe, S\~ao Crist\'ov\~ao, 49100-000, SE, Brazil}

\begin{abstract}

In this letter, we apply the mixed-bond spin-1 Ising model to the study of the magnetic properties of Fe-Mn alloys in the $\alpha$ phase by employing the effective field theory (EFT). Here, we suggest a new approach to the ferromagnetic coupling between nearest neighbours Fe-Fe that depends on the ratio between the Mn-Mn coupling and Fe-Mn coupling and of second power of the Mn concentration $q$ in contrast with linear dependence proposed in the other papers. Also, we propose a new probability distribution for binary alloys with mixed-bonds based on the distribution for ternary alloys and we obtain a very good agreement for all considered values of $q$ in $T-q$ plane, in particular for $q>0.11$.

\end{abstract}


\end{frontmatter}

The effect of diluting a magnetic system by replacing some of the
magnetic atoms with nonmagnetic atoms has attracted the attention of
many researchers and, during the last few years, the theoretical and
experimental study of randomly diluted magnetic systems have contributed to a significant and expressive group of works (see
Refs.~\cite{Barbosa2003,Lara2013,Lima2012,Dias2011,Beutler2005, Mizrahi2004} for more details). In this context, many theoretical problems associated with disordered magnetic
systems at the phase transition have been studied extensively. Some analytical approaches and computational
approximations have been developed in order to treat such systems.
\cite{Wolf2000,Alcazar1986,DosAnjos2007} Among these approaches is
the effective field theory (EFT),~\cite{Honmura1979,Kaneyoshi1981}
based on the identities of Callen and
Suzuki,~\cite{Callen1963,Suzuki1965} that has been used with relative success. In particular, the EFT technique has been applied to
the study of critical phenomena in classical and quantum spin models
which display first and second-order phase transitions as well tricritical points in the phase diagram and has provided useful
qualitative and quantitative insights into the critical behaviour of
these
systems.\cite{Fittipaldi1992,DeAlbuquerque2002,Fittipaldi1992a} These
results have been obtained by treating the effects of the surrounding
spins of each cluster through a convenient differential operator
expansion technique introduced in the literature by Honmura and
Kaneyoshi,\cite{Honmura1979,Kaneyoshi1981} taking all relevant
self-spin correlations into account and including the contribution
of the set of spins.

The Fe-Mn alloys in the $\alpha$ phase ($\alpha$-Fe-Mn) have a bcc
lattice structure\cite{Paduani1991,Paduani1991a,Kulikov1997,Ekholm2011,Mizrahi2004} which is observed
up to about $20$ at. \% Mn. When the Mn concentration $q$
increases, the magnetization linearly decreases up to $q < 0.11$. The average hyperfine field decreases linearly with Mn concentration $q$ up to $20$ at. \% Mn.\cite{Paduani1991,Paduani1991a} The magnetic
properties of the Fe-Mn alloys have been studied extensively by
means of M\"{o}ssbauer effect, nuclear magnetic resonance,
magnetization and others experimental
techniques.\cite{Paduani1991,Paduani1991a,Aeppli1983} Interesting properties of this alloys emerges from other structural phases such as $\gamma$-Fe-Mn (fcc structure) and, in this phase, the Fe-Mn alloys presents antiferromagnetic and glassy behaviour. \cite{Ekholm2011,Aeppli1983} Theoretic and experimental studies shows that the magnetic ground state strongly depends on the lattice parameter and this is a function of the Mn concentration (See Ref. \cite{Ekholm2011} and references therein).

In this letter, we consider the technique usually applied to describe
the magnetic materials that exhibit disorder, namely the diluting
picture and, at the same time, we studied the magnetic properties of
the Fe-Mn alloys, in particular the $T-q$ phase diagram, by
employing EFT. With this in mind, we propose a new probability distribution for binary alloys based on the distribution for ternary alloys.

The outline of the remainder of this paper is as follows: the model
and formalism are described briefly in Section~\ref{section-01}, the
results and discussion are presented in Section~\ref{section-02} and conclusions are presented in Section~\ref{section-03}.

\section{Model and Formalism}\label{section-01}
The Hamiltonian considered for the mixed-bond spin-1 Ising Model is given by
\begin{eqnarray}\label{eq-01}
\mathcal{H} = -\sum_{\langle\,i,\,j\rangle}J_{ij}S_i^z S_j^z\,,
\end{eqnarray}
where the summation is performed over all pairs of the nearest-neighbors sites
 $\langle i,\,j\rangle$ and the quantities $S_i^z$ are
isotropically interacting classical spins localized on the sites $i$
of a bcc  lattice ($S_i^z = \pm 1$ or $0$). In accordance with Pe\~na Lara and co-workers~\cite{PenaLara2009} the exchange interaction $J_{ij}$ obeys the following probability distribution
 
\begin{eqnarray}\label{eq-02}
P(J_{ij}) &=& p^2\delta(J_{ij} - J_1) + q^2\delta(J_{ij} + \gamma J_1) + \nonumber
\\
&+& 2pq\delta(J_{ij} + \lambda J_1),
\end{eqnarray}
where $p^2$ is the probability for bonds ($J_1$) ferromagnetic Fe-Fe atoms, $q^2$ for bonds ($-\gamma\,J_1$) antiferromagnetic Mn-Mn and $2pq$ antiferromagnetic Fe-Mn ($-\lambda J_1$). On the other hand, $\gamma = |J_{MnMn}|/|J_{FeFe}|$ and $\lambda = |J_{FeMn}|/|J_{FeFe}|$ has been used in literature by Paduani and co-workers, \cite{Paduani1991}  where, $p=1-q$ is the Fe concentration, $\gamma = 0.05$ and $\lambda = 0.03$.~\cite{Lara2013,PenaLara2009} Another probability distribution for exchange couplings were proposed for ternary alloys as example Fe-Mn-Al and Fe-Ni-Mn\cite{PenaLara2009,Lara2013,Restrepo2000,Zamora1997} due the asymmetric character of this alloys. The  distribution of the exchange interactions in the Fe-Mn is too asymmetric then, we assume that the distribution is analogous to the ternary alloys with some subtle differences (for example the $2pq$ term for binary alloys in Eq. (\ref{eq-02})) and we use it for the model under consideration.

By employing the EFT with differential operator technique in the one-spin cluster approach, the average magnetization per spin is given by
\begin{eqnarray*}\label{eq-03}
\langle S_i^z\rangle &=&\bigg \langle\prod_je^{K_{ij}S_j^zD_x}\bigg\rangle f(x)\big |_{x=0},
\end{eqnarray*}
where $D_x\equiv\partial/\partial x$, $K_{ij}\equiv \beta \,J_{ij} $ and
$f(x) = \sinh(x) /\left(
\cosh(x) + 1/2\right)\,.$

By employing the generalized van der Waerden identity for spin $S = 1$,~\cite{Tucker1994} one gets:
\begin{eqnarray}\label{eq-04}
\langle S_i^z\rangle &=&\Bigg\langle
\Bigg[ 1+S_j^z \sinh\left(K_{ij}D_x\,\right)\nonumber\\
&+&(S_j^z)^2\,\bigg( \cosh(K_{ij}D_x)-1\,\bigg) \Bigg]^Z\,\Bigg\rangle\,f(x)\bigg|_{x=0}\,,
\end{eqnarray}
where $Z$ is the coordination number ($Z=8$ for bcc lattice). In the vicinity of the second-order phase transition, $m_z \simeq 0\,.$ Then performing the configurational average at the Eq.~(\ref{eq-04}) (here,
denoted by  $m_z =\langle\langle S_i^z\rangle\rangle_c$) and by
expanding up to first order in this parameter, we obtain
\[m_z =A_1(q,\,K_1)\,m_z + \mathcal{O}(m_z^3)\,,\]
with
 \begin{eqnarray}\label{eq-05}
A_1(q,\,K_1) & =& Z\,\Big[(1-q)^2\sinh{(K_1D_x)} - 
q^2\sinh{(\gamma K_1D_x)} \nonumber \\ & - &  2q(1-q)\sinh{(\lambda K_1D_x)} \Big]f(x)\big|_{x=0}\,,
\end{eqnarray}
$K_1\equiv \beta J_1,$ and $A_1(q,\,K_1)$ can be calculated by applying the relation~\cite{Albuquerque2005}
\begin{eqnarray*}
\sinh{(a\,D_x)}\,f(x)\big|_{x=0} = f(a)\,.
\end{eqnarray*}

	In this work, we are interested in the phase boundary of the model under consideration. Then we focus our attention in the second-order transition line, where only the Ising case is studied.~\cite{Kaneyoshi1981,DeAlbuquerque2002}  Since the magnetization $m_z$ goes to zero continuously, a second-order transition line is obtained equation
\begin{eqnarray}
A_{1}(q,\,K_1) = 1\,.\label{T-qline}
\end{eqnarray}

	The Eq. (\ref{eq-05}) shows that the action of the differential operator $\sinh{(aD_x)}$ in $f(x)$ depends on the probability of the each interaction type, $J_1$, $\gamma J_1$ and $\lambda J_1$. The dominant interaction is $J_1 = \beta^{-1} K_1$.  However, in the absence of Fe atoms the $\gamma J_1$ and $\lambda J_1$ interactions determines the sign of the $A_1(q,\,K_1)$ term, and thus the equation for second-order transition line is antiferromagnetic.\cite{Sousa1998}

	In next section, the results and some remarks about the behaviour of the system under consideration are discussed.

\section{Remarks and Discussion}\label{section-02}

\begin{figure}[!htb]
\centering
\includegraphics[scale=0.35,angle=90]{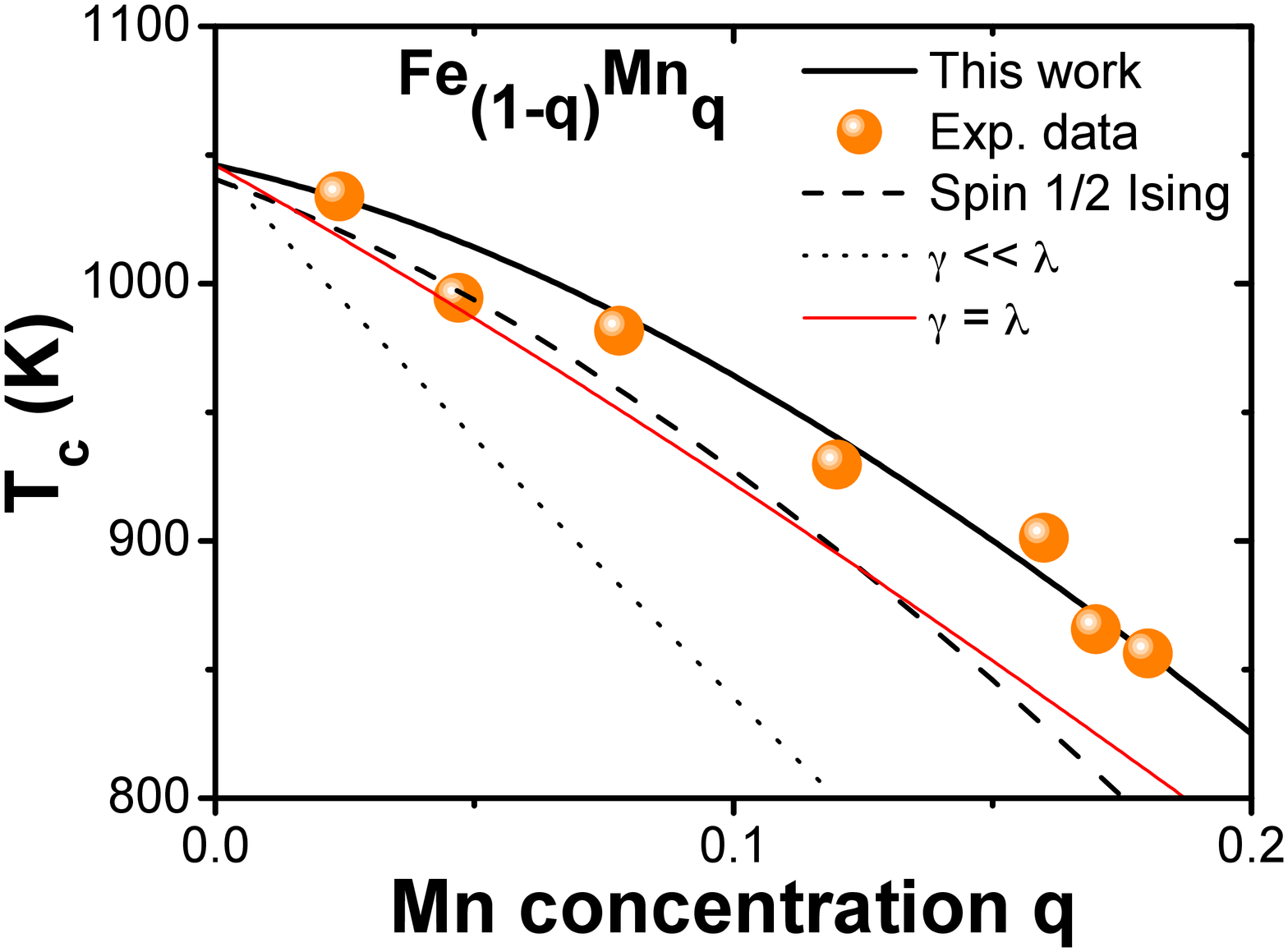}
\caption{Magnetic transition temperature as a function of Mn concentration $q$. The full line is a fit as described in the text, Eq. (\ref{T-qline}), for the spin$-1$ Ising model. Dashed line is the fit via spin$-1/2$ Ising model in approach analogous to Ref.~\cite{Freitas2012b}. Red line represents the case $\gamma = \lambda$ and dotted line $\gamma \ll \lambda$ for spin$-1$ Ising model, respectively. Solid circles are the experimental data, Refs.~\cite{Paduani1991a, Paduani1991, Cabanas2006, Aeppli1983}.}
\label{fig-01}
\end{figure}

	In order to study the $T-q$ phase diagram of disordered Fe-Mn alloys on a bcc lattice, we following the same procedure of Ref.~\cite{Freitas2012b}. The insertion of Mn atoms in alloy produces a variation in the exchange interaction, then we suggest $J_1(q)$ as

\begin{eqnarray}
J_1(q) = J_0\,\Bigl(1 + \frac{\gamma}{\lambda} q(1 - q) \Bigl)\,,\label{jota}
\end{eqnarray}
where $J_0 = 17\,\,meV\,$ is the ferromagnetic coupling for pure iron \cite{Plascak2000,Paduani1991} and $\gamma/\lambda = |J_{MnMn}|/|J_{FeMn}| \approx 1.67$. This result indicates that the Mn-Mn antiferromagnetic coupling is grater than Fe-Mn coupling and, as shown in Fig. \ref{fig-01}, its value significatively influences the agreement of the spin-1 Ising model and experimental results. 

	We can justify the functional dependence of the exchange integral $J_1$ on Mn concentration $q$ by assuming the same behaviour of exchange interaction as in iron-nickel alloys.\cite{Hatherly1964,Krey1978,Kaul1983} Therefore, all exchange interactions in question are not linear. At this point, we consider a numerical treatment which can be done without great difficulty to obtain the phase diagram by using Eq. (\ref{T-qline}), $J_1(q)$ defined by Eq.~(\ref{jota}) and the experimental data.~\cite{Paduani1991a,Paduani1991,Cabanas2006,Aeppli1983} The recursion relation (\ref{T-qline}) provides the critical parameters $K_c^{-1}$ and $q_c$ for the system under consideration. Figure~\ref{fig-01} shows the phase diagram $T-q$ for $0<q<0.2$. It can be seen that there is excellent agreement between experiment and theoretical adjustment throughout all range considered. Thus, it can be seen that the dependence of $T_c$ with Mn concentration is not ``totally'' linear and this suggests that our hypothesis is consistent and we show a comparison with the spin-$1/2$ Ising model in Fig.\ref{fig-01} to demonstrate that the spin$-1$ Ising model improves the theoretical results for considered system, unlike spin-1/2 Ising model which agreement occurs for $q<0.005$. 

	The $\gamma$ and $\lambda$ parameters are relevant to this work and are discussed in Fig. \ref{fig-01}. We observed that for  $\gamma \ll \lambda$ and $\gamma = \lambda$ the $T_c(q)$ linearly decreases when $q$ increases. On the other hand, $dT_c/dq {\rm ( spin-1)} >dT_c/dq {\rm ( spin-1/2)} $ and for spin$-1$ Ising model there is a smooth decrease for $dT_c/dq.$ This result is in agreement with the phenomenological predictions cited in References \cite{Plascak2000,Dias2009,Beutler2005}, in particular for $q<0.2$ indicating that strong ferromagnetic coupling $J_1$ influences the spontaneous magnetization of the Fe-Mn alloys.%

\section{Conclusions}\label{section-03}
	
	We observe that our results are qualitatively and quantitatively consistents with the experimental data for all concentration range of Mn atoms, in contrast to the others approaches in literature carried out using only the linear dependence of the exchange constant as a function of Mn atoms.\cite{Paduani1991,Paduani1991a,Aeppli1983} The mixed-bond spin$-1$ Ising model with nonlinear interactions correctly describes the behaviour of the $T-q$ phase diagram for the Fe-Mn alloys in the $\alpha$ phase and the dependence of the ferromagnetic coupling
between nearest neighbors Fe-Fe with the second power of the Mn
concentration $q$ and allows a simple treatment of the phase diagram
and a good agreement between theory and experiment.

\section*{Acknowledgments}
A. S. Freitas and N. O. Moreno are grateful for the partial support provided by CAPES and CNPq.

\end{document}